# Graphene thermal break-down induced by anharmonic bending mode

V.M. Adamyan[1,†], V.N. Bondarev[1,‡], and V.V. Zavalniuk[1,2,*]

[1] *Department of Theoretical Physics,
Odessa I.I. Mechnikov National University, 2 Dvoryanska St., Odessa 65026, Ukraine*
[2] *Department of Fundamental Sciences,
Odessa Military Academy, 10 Fontanska Road, Odessa 65009, Ukraine*

**Abstract**

The abrupt loss of mechanical stability of two-dimensional graphene-type crystals at a certain transition temperature is described. At this temperature, the graphene state with practically zero-speed bending sound and developed bending fluctuations becomes energetically favorable. Such phenomenon, akin to melting, is naturally caused by the anharmonicity of crystal oscillations. In order to circumvent the known difficulties associated with taking into account the anharmonic effects, we propose an original pseudo-harmonic approximation, within which we determine the free energy of the anharmonic crystal and find a numerical characteristic for the intensity of bending vibrations at transition temperature. This characteristic is similar to the empiric Lindemann criterion for the melting phenomenon. At the same time, in contrast to the conventional Lindemann criterion, the found characteristic is explicitly expressed through the graphene bending moduli of the second, third, and fourth orders.

## I. INTRODUCTION

The creation of a consistent theory of melting is a serious problem of statistical physics [1]. The old phenomenological Lindemann criterion (see, for example, [2, 3]), based on an intuitive connection between the average amplitude of thermal vibrations of a three-dimensional (3D) crystal and its melting point, allowed one to give semi-quantitative interpretation of experimental data for media of various types – from metals to dielectrics [2,3].

However, when attempting to build a direct analog of the Lindemann parameter for graphene as a two-dimensional (2D) crystal of "zero" thickness, the specific difficulties arise, which are absent in 3D solids. For example, the logarithmic divergence of the mean-square thermal fluctuations of atomic displacements for a formally infinite 2D lattice (the Peierls-Landau theorem [4]) makes senseless the direct using the Lindemann criterion in its original form [2] when describing the 2D melting phenomenon. To avoid this, it was proposed [5] to compare with the lattice constant not the absolute root-mean-square fluctuations of atomic displacements (as in the original Lindemann criterion [2]), but the fluctuations of displacements relative to a certain "central" atom. With this approach, the fluctuations really become finite in the limit of an infinite 2D crystal [5], but the result depends on the number of atomic coordination spheres considered in the model, which makes the calculation of the Lindemann 2D criterion ambiguous [6]. Note that the original concept of Ref. 5 is matched to the fundamental principles of the theory of elasticity [7], since both approaches operate with relative displacements of the crystal atoms.

---

[*] vzavalnyuk@onu.edu.ua (corresponding author)
[†] vadamyan@onu.edu.ua
[‡] bondvic@onu.edu.ua

It is significant that the known melting condition of conventional 3D crystals [4], according to which the melting point is determined by the equality of chemical potentials (free energies) of coexisting phases, for graphene as a quasi-2D crystal in 3D space loses meaning [8]. Therefore, as an analog of the graphene "liquid" phase, it was proposed [6,8] to consider regions of quasi-one-dimensional filaments that originate inside the solid graphene when approaching the melting temperature. Besides, in Refs. 6 and 8, using numerical simulation methods, it was shown that at temperatures preceding the melting of graphene, an increase in the number of the so-called Stone-Wales 5775 defects with a formation energy of $\approx 4.6$ eV [9,10] is observed.[1]

Note that in [5] there was no mention of the possible role of the bending mode in the heat transformations of graphene lattice. Until recently, this mode, by analogy with the bending vibrations of thin plates [7], was attributed to the quadratic dispersion law with respect to the wave number $k$ (see, for example, [12]). However, in our work [13] (see also Ref. 14 devoted to the MD simulations of the bending mode), it was shown that the real dispersion law of the bending mode of graphene in the small $k$ region should contain a "sound" – linear in $k$ – segment with anomalously low sound velocity $s_B < 1$ km/s (15–20 times less than the in-plane sound velocities of the graphene layer). An important feature of the graphene bending mode is that its "sound" segment has purely fluctuation origin, therefore the velocity of the bending "sound" is a growing function of temperature [13,15]. On the other hand, the mere fact of the existence of a "sound" dispersion of the bending mode at small $k$ makes the divergence of the out-of-plane mean-square fluctuation displacements in graphene not more "dangerous" than logarithmic [13], which leads to the elimination of the "catastrophic" – power-type on the 2D sample size – divergences resulting from the membrane model [12].

The development of the ideas of paper [13] made it possible to demonstrate [15] the important contribution of the bending acoustic mode to the graphene thermal expansion. The existence of the extended "low-temperature" region of thermal contraction in graphene [16–19] is mainly due to the "soft" bending mode. At the same time, the effect of the graphene thermal contraction, itself, was consistently described [15] only by going beyond the quasi-harmonic approximation.

In this paper, we show that like in [13,15] the anomalously "soft" bending mode play a major role in the phase transition that can be interpreted as an analog of graphene melting. We will see that due to the specificity of this mode at a certain temperature graphene jumps to the state with anomalously large amplitude of out-of-plane fluctuation displacements (Section II), which can be considered as a sign of melting, whatever the nature of the emerging "melt".[2] Hereinafter, for definiteness, the corresponding phase transition is conventionally called "melting". The key point in the very possibility of such a transition, as it turns out, is negativity of the anharmonic third-order elastic moduli of graphene. The modeling of graphene as 2D elastic continuum with certain elastic coefficients allows us not only to introduce the concept of bending "sound" in a consistent way, but also to find a convergent value for the 2D analog of Lindemann criterion (Section III). The numerical results given in Section IV demonstrate the realistic values of characteristic parameters (the melting temperature, the out-of-plane and in-plane mean-square fluctuations of the graphene strain tensor) expressed through the assumingly known elastic moduli of the second-, third-, and fourth-order. In Section V, conclusions are formulated and ways for further research into the problem of quasi-2D melting are outlined. Some intermediate calculations and considerations are included into Appendices A and B.

---

[1] It is worth mentioning the arguments given in [8], according to which the melting of graphene can hardly occur by the mechanism proposed in [11].

[2] Note that an attempt to study the melting of graphene by MD simulations in the framework of non-realistic model of strictly 2D crystal led to the melting temperature $\approx 8000$ K [20]. The latter value is much greater than the value $\approx 4500$ K [6,8] that usually follows from numerical simulations of graphene melting with considering the bending degree of freedom. This result demonstrates the crucial role of the bending mode in the graphene melting.

## II. ANHARMONIC EFFECTS AND BASIC PRINCIPLES OF THE THEORY OF GRAPHENE MELTING

As in our previous studies [13,15], we simulate graphene as a quasi-2D elastic medium described by quasi-2D local (dependent on time $t$) strain tensor $\varepsilon_{\alpha\beta}(\mathbf{r},t)$ [7]:

$$\varepsilon_{\alpha\beta}(\mathbf{r},t) = \frac{1}{2}[\partial_\alpha u_\beta(\mathbf{r},t) + \partial_\beta u_\alpha(\mathbf{r},t) + \partial_\alpha u_\gamma(\mathbf{r},t)\partial_\beta u_\gamma(\mathbf{r},t) + \partial_\alpha w(\mathbf{r},t)\partial_\beta w(\mathbf{r},t)], \quad \partial_\alpha \equiv \frac{\partial}{\partial r_\alpha}. \quad (1)$$

Here $\mathbf{u}(\mathbf{r},t)$, with $\mathbf{r}=(x,y)$, is the 2D displacement vector of the elastic medium along the unstrained graphene plane, $w(\mathbf{r},t)$ is the corresponding out-of-plane displacement, Greek indices run over two values, $x$ and $y$, and summation is assumed over repeated indices. Immediately, we note that the products of the derivatives in (1) are not discarded anywhere further.

Having in mind to describe the melting of graphene as a substantially nonlinear effect we represent the mechanical Hamiltonian of graphene taking into account the anharmonic terms of the third [13,15] and fourth orders with respect to $\varepsilon_{\alpha\beta}$ (in what follows we omit the arguments of the field functions):

$$\begin{aligned}\boldsymbol{H}\{\mathbf{u},w\} = \int d\mathbf{r}\Bigg[&\frac{\rho}{2}(\dot{\mathbf{u}}^2+\dot{w}^2) + \frac{\lambda}{2}\varepsilon_{\alpha\alpha}\varepsilon_{\beta\beta} + \mu\varepsilon_{\alpha\beta}\varepsilon_{\alpha\beta} + \frac{\kappa}{2}(\nabla^2 w)^2 \\ &+ \frac{C_{111}-C_{112}}{4}\varepsilon_{\alpha\beta}\varepsilon_{\alpha\beta}\varepsilon_{\gamma\gamma} + \frac{3C_{112}-C_{111}}{12}\varepsilon_{\alpha\alpha}\varepsilon_{\beta\beta}\varepsilon_{\gamma\gamma} \\ &+ \frac{1}{4!}\Big(a_4\varepsilon_{\alpha\beta}\varepsilon_{\beta\gamma}\varepsilon_{\gamma\delta}\varepsilon_{\delta\alpha} + b_4\varepsilon_{\alpha\beta}\varepsilon_{\alpha\beta}\varepsilon_{\gamma\gamma}\varepsilon_{\delta\delta} + c_4\varepsilon_{\alpha\beta}\varepsilon_{\alpha\beta}\varepsilon_{\gamma\delta}\varepsilon_{\gamma\delta} + d_4\varepsilon_{\alpha\alpha}\varepsilon_{\beta\beta}\varepsilon_{\gamma\gamma}\varepsilon_{\delta\delta}\Big)\Bigg].\end{aligned} \quad (2)$$

Here $\rho$ is the 2D mass density of graphene, $\mu > 0$ and $\lambda > -\mu$ (in fact, $\lambda > 0$, see [7,21]) are the 2D Lamé coefficients; $C_{111} < 0$ and $C_{112} < 0$ [22–24] are the third order elastic coefficients (we put $C_{222} = C_{111}$, as in isotropic model) whereas $a_4$, $b_4$, $c_4$, and $d_4$ are the fourth order elastic coefficients. Besides, $\kappa$ is the so-called bending rigidity and $\nabla$ is the 2D gradient; $\dot{\mathbf{u}}$ and $\dot{w}$ denote the so-called material time-derivatives [25] of the local displacements of material points of graphene as a quasi-2D continuum medium (for more details, see Ref. 15).

From (2) with taking account of (1), it follows that the quantized fields $\mathbf{u}$ and $w$ are formally interacting. However, as shown by detailed analysis in [13,15], their interaction in graphene is effectively eliminated at each temperature virtually by heat averaging the cross-components like $\partial_\alpha u_\gamma \partial_\beta u_\gamma \partial_\alpha w \partial_\beta w$ and $\partial_\alpha w \partial_\beta w(\partial u_\alpha/\partial t)(\partial u_\beta/\partial t)$ in (2) using the density operator for the harmonic Hamiltonian $\boldsymbol{H}_\mathbf{u}$ of the "fast" in-plane modes:

$$\boldsymbol{H}_\mathbf{u} = \int d\mathbf{r}\left[\frac{\rho}{2}\left(\frac{\partial \mathbf{u}}{\partial t}\right)^2 + \frac{\lambda}{2}u_{\alpha\alpha}u_{\beta\beta} + \mu u_{\alpha\beta}u_{\alpha\beta}\right], \quad (3)$$

where $u_{\alpha\beta} \equiv (\partial_\alpha u_\beta + \partial_\beta u_\alpha)/2$ is the linear part of the strain tensor (1).

In this way we get, first of all, the effective harmonic [superscript (h)] Hamiltonian for the bending mode of graphene [13]:

$$\boldsymbol{H}_w^{(\mathrm{h})} = \int d\mathbf{r}\left[\frac{\rho}{2}\left(\frac{\partial w}{\partial t}\right)^2 + \frac{B(T)}{2}(\nabla w)^2 + \frac{\kappa}{2}(\nabla^2 w)^2\right]. \quad (4)$$

Here, the temperature-dependent bending modulus of graphene [13,15]

$$B(T) = \frac{1}{4}\left(3\lambda + 5\mu + \frac{3C_{111} + C_{112}}{4}\right)\langle \partial_\alpha u_\beta \partial_\alpha u_\beta \rangle_{\mathbf{u}} \qquad (5)$$

has purely fluctuation nature being proportional to the thermal average $\langle ... \rangle_{\mathbf{u}}$ with the density operator defined by (3). The dispersion law of the bending mode following from (4)

$$\omega_B^2(k) = s_B^2 k^2 + \frac{\kappa}{\rho} k^4, \quad s_B = \sqrt{\frac{B(T)}{\rho}} \qquad (6)$$

is of the acoustic type with the sound velocity $s_B$ at small wave numbers [13,15]. Namely this "soft" bending mode may be considered as a factor due to which graphene can most easily lose its stability at high temperatures.

The explicit expression for the thermal average in (5) found within the framework of the Debye model [4] is [13,15]:

$$\langle \partial_\alpha u_\beta \partial_\alpha u_\beta \rangle_{\mathbf{u}} = \frac{(\lambda + 3\mu)\theta_\| \rho}{3m\mu(\lambda + 2\mu)}\left[1 + 6\left(\frac{T}{\theta_\|}\right)^3 \int_0^{\theta_\|/T} \frac{\xi^2 d\xi}{e^\xi - 1}\right], \qquad (7)$$

where the atomic mass $m$ appears (in the case of graphene it is the mass of $^{12}C$) and the in-plane Debye temperature is

$$\theta_\| = 2\hbar\sqrt{\frac{2\pi \mu(\lambda + 2\mu)}{m(\lambda + 3\mu)}} \approx 2670 \text{ K}. \qquad (8)$$

In the formal limit $T \to \infty$ (we will need it below) from (7) we obtain (see [13]):

$$\langle \partial_\alpha u_\beta \partial_\alpha u_\beta \rangle_{\mathbf{u}} = \frac{\rho(\lambda + 3\mu)}{m\mu(\lambda + 2\mu)} T. \qquad (9)$$

Expression (4) for the Hamiltonian of the bending mode is written in the harmonic approximation. Meanwhile, already from (2) and representation (1) one can establish the following structure of the fourth, sixth, and eighth power terms in the density of the Hamiltonian:

$$\frac{\lambda + 2\mu}{8}(\nabla w)^4 + \frac{C_{111}}{48}(\nabla w)^6 + \frac{a_4 + b_4 + c_4 + d_4}{384}(\nabla w)^8. \qquad (10)$$

Immediately, we note that the form (10) arises due to the specificity of a quasi-2D crystal, which allows for the presence in the strain tensor (1) only terms, quadratic in the space derivatives of the out-of-plane displacements of graphene.

As noted above, both experiments [23] and simulations [22,24,26] demonstrate that the third-order elastic modulus $C_{111}$ in graphene is negative. This means that just the last term in (10) with a positive coefficient determined by fourth-order elastic moduli in Hamiltonian (2) ensures the stability of graphene with respect to large bending fluctuations.

The question of the structure of the 3rd, 4th, and 5th order anharmonic contributions to the elastic in-plane Hamiltonian of graphene was considered in [24], where the results of [22] for $C_{111} < 0$ and $C_{112} < 0$ were confirmed by numerical simulation, and also numerical estimates were given for the elastic moduli of the fourth and fifth orders. In this case, according to [24], the fourth-order moduli for graphene turned out to be positive, and the fifth-order ones – negative. Therefore, adding the density of the Hamiltonian (4) to the anharmonic terms of the form (10), we obtain the resulting expression for the "bending" Hamiltonian of graphene in the form:

$$\boldsymbol{H}_w = \int d\mathbf{r} \left[ \frac{\rho}{2}\left(\frac{\partial w}{\partial t}\right)^2 + \frac{B(T)}{2}(\nabla w)^2 + \frac{\kappa}{2}(\nabla^2 w)^2 \right.$$
$$\left. + \frac{C_{11}(T)}{2!2^2}(\nabla w)^4 + \frac{C_{111}}{3!2^3}(\nabla w)^6 + \frac{C_{1111}}{4!2^4}(\nabla w)^8 \right]. \tag{11}$$

This expression contains a "consolidated" fourth-order elastic modulus $C_{1111} \equiv a_4 + b_4 + c_4 + d_4 > 0$,[3] the value of which, in principle, can be reconstructed from the results of numerical simulations for graphene [24,26]. As for the factor $C_{11}(T)$ at $(\nabla w)^4$ in (11), then, as a generalization of (10), we allowed the possibility of temperature dependence of the second-order elastic modulus, with $C_{11}(0) = \lambda + 2\mu$ [below an explicit expression for $C_{11}(T)$ will be given]. As for the "anharmonic" elastic moduli $C_{111} < 0$ and $C_{1111} > 0$, then, if we not go beyond the goals of this work, they can be considered constants.

A direct calculation of the contribution $F_w(T)$ based on the Hamiltonian (11) to the total free energy of graphene so far is not real. We have to some extent overcome this difficulty replacing approximately $\boldsymbol{H}_w$ by some pseudo-harmonic [superscript (p-h)] "bending" Hamiltonian[4]:

$$\boldsymbol{H}_w^{(\text{p-h})} = \int d\mathbf{r} \left\{ \frac{\rho}{2}\left(\frac{\partial w}{\partial t}\right)^2 + \frac{\rho}{2}[\hat{\omega}(\nabla;T)w]^2 \right\}, \tag{12}$$

which will define the pseudo-harmonic "bending" free energy $F_w^{(\text{p-h})}(T)$ (see below). The pseudo-differential operator $\hat{\omega}(\nabla;T)$ appearing in (12), depends on temperature as a parameter and is defined in the following way. We write the direct and inverse 2D Fourier transformations of the graphene bending displacement (we omit the temporal argument):

$$w_{\mathbf{k}} = \frac{1}{\sqrt{S}} \int e^{-i\mathbf{k}\mathbf{r}} w(\mathbf{r}) d\mathbf{r}, \quad w(\mathbf{r}) = \frac{1}{\sqrt{S}} \sum_{\mathbf{k}} w_{\mathbf{k}} e^{i\mathbf{k}\mathbf{r}} \quad w_{-\mathbf{k}} = w_{\mathbf{k}}^*, \tag{13}$$

where $S$ is the graphene sheet area. We require that the action of the operator $\hat{\omega}(\nabla;T)$ on the function $w(\mathbf{r})$ leads in the Fourier representation to the expression:

$$\omega_B^{(\text{p-h})}(k) w_{\mathbf{k}} = \frac{1}{\sqrt{S}} \int e^{-i\mathbf{k}\mathbf{r}} \hat{\omega}(\nabla;T) w(\mathbf{r}) d\mathbf{r}. \tag{14}$$

---

[3] Hereinafter, we denote this "consolidated" modulus as $C_{1111}$ and further use the concept of isotropic graphene, in which $C_{1111} = C_{2222}$ (cf. the above similar equality $C_{222} = C_{111}$). In this case, the rest of 4th order elastic constants will be assumed small in comparison with $C_{1111}$ [24].

[4] It is worth noting that the term "pseudo-harmonic approximation" was used in [27] when considering the problem of self-consistent oscillations of an anharmonic lattice.

In this case, the desired energy $\hbar\omega_B^{(p-h)}(k)$ of the bending vibration excitation with the wave number $k$ should be chosen so that at a given temperature $T$ the free energy $F_w^{(p-h)}(T)$ of the model system with Hamiltonian (12) is the least deviated from the "true" free energy $F_w(T)$ of the subsystem of bending excitations with Hamiltonian (11).

It is convenient to search for an explicit expression for $\omega_B^{(p-h)}(k)$, based on a theorem first established by Peierls [28] (see also [29,30]). In terms of the quantities introduced above, the theorem [28] reads as follows:

$$F_w(T) \leq F_w^{(p-h)}(T) + \langle \boldsymbol{H}_w - \boldsymbol{H}_w^{(p-h)} \rangle_w. \tag{15}$$

We will consider the right-hand side of inequality (15) as a functional of variable $E_\mathbf{k} \equiv \hbar\omega_B^{(p-h)}(k)$ and find its minimum by this variable using expressions (11)–(14).

Representing the "pseudo-harmonic" free energy $F_w^{(p-h)}(T)$ in explicit form [2], one can write the right-hand side of (15) as a functional of $E_\mathbf{k}$:

$$J\{E_\mathbf{k}\} \equiv \sum_\mathbf{k} \left[ \frac{E_\mathbf{k}}{2} + T\ln\left(1 - e^{-E_\mathbf{k}/T}\right) \right] + \langle \boldsymbol{H}_w - \boldsymbol{H}_w^{(p-h)} \rangle_w. \tag{16}$$

Further, when calculating the averages in (16), we take into account that

$$\left\langle \int (\nabla w)^{2n} d\mathbf{r} \right\rangle_w = \int \left\langle (\nabla w)^{2n} \right\rangle_w d\mathbf{r} = S\left\langle (\nabla w)^{2n} \right\rangle_w, \quad n \geq 1, \tag{17}$$

and then instead of (16) we get:

$$J\{E_\mathbf{k}\} = T\sum_\mathbf{k} \ln\left[2\sinh\left(\frac{E_\mathbf{k}}{2T}\right)\right] + \frac{1}{2}\sum_\mathbf{k}\left[B(T)k^2 + \kappa k^4 - \rho\frac{E_\mathbf{k}^2}{\hbar^2}\right]\langle |w_\mathbf{k}|^2 \rangle_w$$
$$+ S\left[\frac{C_{11}(T)}{2!2^2}\langle (\nabla w)^4 \rangle_w + \frac{C_{111}}{3!2^3}\langle (\nabla w)^6 \rangle_w + \frac{C_{1111}}{4!2^4}\langle (\nabla w)^8 \rangle_w\right]. \tag{18}$$

Since the averaging in (18) is carried out with the pseudo-harmonic Hamiltonian (12), the following formula takes place (for details see Appendix A):

$$\langle (\nabla w)^{2n} \rangle_w = n!\langle (\nabla w)^2 \rangle_w^n, \quad n \geq 1. \tag{19}$$

Thus, the averages appearing in (18) can be expressed in terms of the only quantity:

$$q \equiv \frac{1}{2}\langle (\nabla w)^2 \rangle_w = \frac{1}{2S}\sum_\mathbf{k} k^2 \langle |w_\mathbf{k}|^2 \rangle_w, \tag{20}$$

where (see [15])

$$\langle |w_\mathbf{k}|^2 \rangle_w = \frac{\hbar^2}{2\rho E_\mathbf{k}}\coth\left(\frac{E_\mathbf{k}}{2T}\right). \tag{21}$$

Now, taking into account (19)–(21), we reduce (18) to the form:

$$J\{E_{\mathbf{k}}\} = T\sum_{\mathbf{k}} \ln\left[2\sinh\left(\frac{E_{\mathbf{k}}}{2T}\right)\right] + \frac{1}{4}\sum_{\mathbf{k}}\left\{\frac{\hbar^2}{\rho E_{\mathbf{k}}}[B(T)k^2 + \kappa k^4] - E_{\mathbf{k}}\right\}\coth\left(\frac{E_{\mathbf{k}}}{2T}\right)$$
$$+ S[C_{11}(T)q^2 + C_{111}q^3 + C_{1111}q^4]. \tag{22}$$

Varying (22) over $E_{\mathbf{k}}$ [with taking account of the variation of $q$ by Eqs. (20) and (21)] and equating the result to zero, we arrive at the equation:

$$\frac{\delta J\{E_{\mathbf{k}}\}}{\delta E_{\mathbf{k}}} = \frac{1}{4}\coth\left(\frac{E_{\mathbf{k}}}{2T}\right)\left[1 + \frac{E_{\mathbf{k}}}{T\sinh(E_{\mathbf{k}}/T)}\right]$$
$$\times\left\{1 - \frac{\hbar^2 k^2}{\rho E_{\mathbf{k}}^2}[B(T) + 2C_{11}(T)q + 3C_{111}q^2 + 4C_{1111}q^3 + \kappa k^2]\right\} = 0, \tag{23}$$

which provides a minimum (more exactly, an extremum) of the functional $J\{E_{\mathbf{k}}\}$. As a result, instead of expression (6), we obtain from (23) the dispersion equation for the bending mode in the framework of the "pseudo-harmonic" problem:

$$\rho[\omega_B^{(\text{p-h})}(k)]^2 = [B(T) + 2C_{11}(T)q + 3C_{111}q^2 + 4C_{1111}q^3]k^2 + \kappa k^4, \tag{24}$$

where the parameter $q$ defined by (20) contains integrated information about the spectrum $\omega_B^{(\text{p-h})}(k) \equiv E_{\mathbf{k}}/\hbar$, and, of course, the coefficient at $k^2$ in (24) is assumed to be positive. More sophisticated consideration shows that the extremals of $J\{E_{\mathbf{k}}\}$ obtained in this way will give precisely the minima if the quantities found from (20) and (24) satisfy the additional condition

$$1 + \frac{1}{4\rho^2 S}[2C_{11}(T) + 6C_{111}q + 12C_{1111}q^2]\sum_{\mathbf{k}}\coth\left(\frac{E_{\mathbf{k}}}{2T}\right)\left[1 + \frac{E_{\mathbf{k}}}{T\sinh(E_{\mathbf{k}}/T)}\right]\frac{\hbar^4 k^4}{E_{\mathbf{k}}^3} > 0,$$

which is definitely satisfied in our task.

Here, however, it is necessary to dwell on the question of the temperature range of applicability of the "pseudo-harmonic" approximation (24). We show that at "low" temperatures (we will later specify this definition), taking into account the first term in (10) will only affect the coefficient at $k^4$ in the dispersion equation (6), but will not affect the velocity of bending sound $s_B$. It is easy to verify that when taking into account the mentioned anharmonic term, the equation of motion for the bending mode will take the form:

$$\frac{\partial^2 w}{\partial t^2} = s_B^2 \partial_{\alpha\alpha}^2 w - \frac{\kappa}{\rho}\partial_{\alpha\alpha\beta\beta}^4 w + \frac{\lambda + 2\mu}{2\rho}(\partial_{\alpha\alpha}^2 w \partial_\beta w \partial_\beta w + 2\partial_\alpha w \partial_\beta w \partial_{\alpha\beta}^2 w). \tag{25}$$

Suppose now that in a graphene sample there is a single bending wave with the wave vector $\mathbf{k}$, frequency $\omega$, and amplitude $w_{\mathbf{k}}$:

$$w(\mathbf{r},t) = w_{\mathbf{k}}\cos(\mathbf{k}\mathbf{r} - \omega t). \tag{26}$$

Substituting (26) into (25) and omitting terms with triple harmonics, we obtain instead of (6) the dispersion law of the bending mode with taking account of the nonlinear contribution (self-action):

$$\omega_B^2(k) = s_B^2 k^2 + \frac{\kappa}{\rho} k^4 + \frac{3(\lambda + 2\mu)}{16\rho} w_{\mathbf{k}}^2 k^4. \qquad (27)$$

Thus, the inclusion of anharmonic terms in the "bending" Hamiltonian of graphene does not lead to the modification of the long-wave – "sound" – part of the bending mode spectrum at low temperatures. However, the role of the last term in (27) will increase with increasing temperature due to the natural growth of the mean square amplitude of the graphene bending mode. On the other hand, at finite temperatures, harmonics with different (including large) wave numbers are excited in graphene. As a result, the anharmonic terms of the form [cf. (27)]

$$\sim \frac{\lambda + 2\mu}{\rho} k^2 \sum_{\mathbf{k}_1 \neq \mathbf{k}} k_1^2 w_{\mathbf{k}_1}^2 \qquad (28)$$

appear in the dispersion equation of the bending mode; at that, the short-wave harmonics give the main contribution to the sum in (28). The terms of the form (28) after thermodynamic averaging of the sum will lead to the modification of the "sound" part of the graphene bending vibrations, which, in fact, is reflected in equation (24). Such a modification, however, can effectively manifest itself only at sufficiently high temperatures.[5]

The scale of the temperature, above which expressions (12) and (24) take effect, is naturally associated with the out-of-plane Debye temperature (see [15]):

$$\theta_w \equiv \frac{4\pi\hbar}{m} \sqrt{\kappa\rho} = 2040 \text{ K}. \qquad (29)$$

Interestingly, at a temperature $T_0 \approx \theta_w/2 \approx 1000$ K, the thermal expansion coefficient of graphene changes from negative to positive values [15,17,18]. Then it can be expected that in the limit of "low" temperatures ($T \ll T_0$), the dispersion law (6) with the "bare" bending modulus $B(T)$ should take place, whereas at high temperatures, the full expression (24) "works". Therefore, one can propose the following simple interpolation formula for a "pseudo-harmonic" bending modulus:

$$B(Q;T) \equiv B(T) + B_0(T)[\eta(T)Q - 3Q^2 + 3Q^3], \quad B_0(T) \equiv \frac{9|C_{111}|^3}{16 C_{1111}^2} \frac{T}{T+T_0}. \qquad (30)$$

In formula (30), the redefined mean square of the gradient of the graphene out-of-plane displacement is presented

$$Q \equiv \frac{2 C_{1111}}{3|C_{111}|} \left\langle (\nabla w)^2 \right\rangle_w; \qquad (31)$$

---

[5] The reasoning given here resembles the arguments of work [31], in which the attenuation of the high-intensity sound in a lined acoustic duct was studied using the harmonic linearization method [32,33]. It is useful to note that, as shown in [13], the contribution of $\sim k^4$ to the dispersion law of the bending mode due to the term $\sim (\lambda + 2\mu)(\nabla w)^4$ also takes place at finite temperatures. This, in fact, means that the terms of the form of (28), which have anharmonic origin, can manifest themselves only in the high-temperature limit. This question, however, requires more study.

in addition, it was taken into account that $C_{111} < 0$ [22–24], and a dimensionless "controlling" parameter [dependent on temperature through $C_{11}(T)$] was introduced

$$\eta(T) \equiv \frac{8C_{1111}}{3|C_{111}|^2} C_{11}(T). \tag{32}$$

Moreover, from (12) – (14) we find the desired "pseudo-harmonic" Hamiltonian:

$$\boldsymbol{H}_w^{(\text{p-h})} = \int d\mathbf{r} \left[ \frac{\rho}{2}\left(\frac{\partial w}{\partial t}\right)^2 + \frac{1}{2} B(Q;T)(\nabla w)^2 + \frac{\kappa}{2}(\nabla^2 w)^2 \right]. \tag{33}$$

Accordingly, instead of (6), we obtain the dispersion law of the "pseudo-harmonic" bending mode of graphene:

$$[\omega_B^{(\text{p-h})}(k)]^2 = s_B^2(Q;T)k^2 + \frac{\kappa}{\rho}k^4, \quad s_B(Q;T) \equiv \sqrt{\frac{B(Q;T)}{\rho}}. \tag{34}$$

## III. THRESHOLD SINGULARITY OF THE BENDING MODE AS A SIGN OF GRAPHENE MELTING

If $\eta(T) < 1$, then, as is easily seen from (30), function $B(Q;T)$ passes through a maximum and a minimum at points $Q_-(T) = [1 - \sqrt{1-\eta(T)}]/3$ and $Q_+(T) = [1 + \sqrt{1-\eta(T)}]/3$, respectively. Further, if $\eta(T) < 3/4$, then function $[\eta(T)Q - 3Q^2 + 3Q^3]$ has a range of negative values. The latter circumstance is especially important since in this case it is possible to achieve small (in principle, arbitrarily small) values of the "pseudo-harmonic" bending modulus, which is a necessary prerequisite for the implementation of solutions with "large" (~ 1) values of $Q$. The abrupt transition from $Q \ll 1$ to $Q \sim 1$ upon reaching a certain temperature $T_m$ will be interpreted as a sign of the graphene melting.

It is appropriate mention here that the use of the function $[\eta(T)Q - 3Q^2 + 3Q^3]$, which depends on the redefined dimensionless variable $Q$, makes it possible to describe the melting phenomenon in the most general form, not limited to the specific numerical values of the elastic moduli of different orders. We will see that the appearance of a "special" (with $Q \sim 1$) solution of a self-consistent equation is of a fundamentally *threshold* nature. In this case, at the very threshold point the solution with $Q \sim 1$ will correspond to a lower free energy of the graphene bending mode fluctuations than the "regular" (with $Q \ll 1$) solution.

In the framework of the proposed approach, the analysis of the phenomenon of graphene melting can be given, based on the expression for $\langle (\nabla w)^2 \rangle_w$, derived in our works [13,15]. Replacing the "bare" bending modulus $B(T)$ by the "pseudo-harmonic" one $B(Q;T)$ in accordance with (30), (33) and taking into account (31) and (34), we arrive at a self-consistent equation for finding possible values of $Q$:

$$Q = \frac{\hbar C_{1111}}{6\pi\rho |C_{111}|} \int_0^{k_{\max}} \frac{k^3 dk}{\omega_B^{(\text{p-h})}(k)} \left[ \frac{1}{e^{\hbar \omega_B^{(\text{p-h})}(k)/T} - 1} + \frac{1}{2} \right], \tag{35}$$

where $k_{max} = \sqrt{4\pi\rho/m}$ is the maximum wave-number in the Debye model [13,15]. Further, since the melting of graphene occurs at a temperature of ≈ 4500 K, the right-hand side of (35) can be taken in the high temperature limit (as in Ref. 13), and then we obtain the desired self-consistent equation in the form:

$$Q = \frac{C_{1111}T}{6\pi|C_{111}|\kappa}\ln\left[1 + \frac{4\pi\kappa\rho}{mB(Q;T)}\right]. \tag{36}$$

First of all, it is obvious that Eq. (36) at all temperatures contains the above mentioned "regular" (for details see below) solution, $Q^{(R)} \ll 1$ [superscript (R)]. In addition, using Eq. (36), one can determine the threshold for the appearance of the desired "special" solution $Q^{(S)} \sim 1$ [superscript (S)]. To do this, it suffices to find the temperature $T_m$ and the value of variable $Q_m^{(S)}$ at the point of tangency of the left-hand and right-hand sides of Eq. (36). One of the conditions of tangency will be the fulfillment of equality (36) itself, and the second is the equality of the derivatives of the left and right sides of (36) at $T = T_m$ and $Q = Q_m^{(S)}$. As a result, two equations for determining $T_m$ and $Q_m^{(S)}$ will have the form:

$$Q_m^{(S)} = \frac{C_{1111}T_m}{6\pi|C_{111}|\kappa}\ln\left[1 + \frac{4\pi\kappa\rho}{mB(Q_m^{(S)};T_m)}\right], \tag{37}$$

$$1 = -\frac{2C_{1111}T_m\rho B'(Q_m^{(S)};T_m)}{3|C_{111}|[4\pi\kappa\rho + mB(Q_m^{(S)};T_m)]B(Q_m^{(S)};T_m)}, \tag{38}$$

where

$$B'(Q_m^{(S)};T_m) \equiv \left.\frac{\partial B(Q;T_m)}{\partial Q}\right|_{Q_m^{(S)}}. \tag{39}$$

The temperature $T_m$ at which the "special" solution $Q_m^{(S)}$ abruptly appears could be naturally identified with the melting temperature of graphene.

## IV. SELF-CONSISTENT CALCULATION OF THE GRAPHENE MELTING CRITERION AND NUMERICAL ESTIMATES

The solution of system (37), (38) is radically simplified due to the following circumstance. As will be shown, the required value $Q_m^{(S)} \approx 0.5$ is many times greater than the coefficient in front of the logarithm on the right-hand side of Eq. (37). Indeed, taking for estimation $C_{1111} \approx |C_{111}|$ (we will see this below) and assuming $\kappa \approx 1.5$ eV [21], $T_m \approx 4500$ K [6,8], we have $3\pi|C_{111}|\kappa/(C_{1111}T_m) \approx 36$, i.e. $mB(Q_m^{(S)};T_m)/(4\pi\kappa\rho) \sim 10^{-17}$. This means that in Eq. (38) with the highest accuracy one can put $B(Q_m^{(S)};T_m) = 0$, and therefore $B'(Q_m^{(S)};T_m) = 0$, as well. From here we get two equalities:

$$\eta(T_m)Q_m^{(S)} - 3[Q_m^{(S)}]^2 + 3[Q_m^{(S)}]^3 = -\frac{B(T_m)}{B_0(T_m)}, \quad \eta(T_m) - 6Q_m^{(S)} + 9[Q_m^{(S)}]^2 = 0. \tag{40}$$

Thus, one of the conditions for the appearance of the "special" solution of equation (36) is the practical vanishing of the "pseudo-harmonic" bending modulus at the threshold point. This result, in some sense, returns us to the old idea [34] that when approaching the melting point of a crystal, its shear modulus should tend to zero. However, it was later shown [3] that the assumption [34] is inconsistent with the state of affairs in the melting of real crystals. However, using the example of graphene, we have seen that the "pseudo-harmonic" bending modulus as some effective combination of the elastic moduli (including anharmonic ones!) really undergo an extreme softening at the melting point of the material.

If $B(T_\mathrm{m})/B_0(T_\mathrm{m}) \ll 1$ (and this is true for graphene, see below), then up to the principal terms, the solutions of Eqs. (40) are as follows:

$$Q_\mathrm{m}^{(S)} = \frac{1}{2} + \frac{2B(T_\mathrm{m})}{3B_0(T_\mathrm{m})}, \quad \eta(T_\mathrm{m}) = \frac{3}{4} - \frac{2B(T_\mathrm{m})}{B_0(T_\mathrm{m})}. \qquad (41)$$

Substituting the solution for $\eta(T_\mathrm{m})$ from (41) into (32), we immediately obtain an expression relating the elastic moduli of different orders between themselves at the melting point of graphene:

$$\frac{32 C_{1111}}{9|C_{111}|^2} C_{11}(T_\mathrm{m}) + \frac{8B(T_\mathrm{m})}{3B_0(T_\mathrm{m})} = 1. \qquad (42)$$

Eq. (42) implicitly determines the melting temperature $T_\mathrm{m}$ of graphene, which can be considered as one of the central results of the paper. If we neglect the relatively small second term on the left-hand side of Eq. (42), then to find $T_\mathrm{m}$, it is necessary to have information about the temperature dependence of the second-order elastic modulus $C_{11}(T)$. Then, with the known values of the anharmonic constants $C_{111}$ and $C_{1111}$, the value of $T_\mathrm{m}$ for graphene can be found from Eq. (42).

The dependence $C_{11}(T)$ can be established using the results of simulations [35], wherefrom one can determine the value of $C_{11}(T_\mathrm{m})$. Taking $C_{11}(0) \approx 330$ N/m [17], estimating the change of $C_{11}(T)$ with temperature according to [35] and assuming $T_\mathrm{m} \approx 4500$ K [6,8], we get $C_{11}(T_\mathrm{m}) \approx 293$ N/m, which corresponds to a relative change $[C_{11}(0) - C_{11}(T_\mathrm{m})]/C_{11}(0) \approx 11\%$. However, given below in this section (see also Appendix C) our estimate of such a change is about half as much, which gives $C_{11}(T_\mathrm{m}) \approx 314$ N/m.

Equation (42) is especially important because, as far as the authors know, only the paper [24] reported the calculation of the elastic constants of the fourth (and also the fifth!) order for graphene using the density functional theory (DFT) method. In paper [15], to match our results on the velocity of the bending "sound" with the simulation data [14], we obtained the estimate $C_{111} \approx -1000$ N/m. Then from Eq. (42) we find our estimate: $C_{1111} \approx 900$ N/m (what about the comparison with the results of [24], see below).

On the other hand, information about fourth-order elastic moduli for graphene can be obtained by analyzing the results of [26] obtained by MD simulations and DFT calculations. As for third-order elastic moduli (we assume $C_{111} = C_{222}$), it was actually established in [26] that their possible values in graphene lie in the range from −1100 N/m to −1400 N/m (cf. the above estimate from [15]). However, the question of fourth order moduli was not discussed in [26]. Meanwhile, based on the Hamiltonian of graphene (2) and considering the uniaxial strain $\varepsilon$, we can write the following non-linear expression for uniaxial stress (cf. [24]):

$$\sigma = C_{11}(0)\varepsilon + \frac{1}{2}C_{111}\varepsilon^2 + \frac{1}{6}C_{1111}\varepsilon^3, \tag{43}$$

where we put $T = 0$, referring to data analysis [26], and also retained the notation $C_{1111}$ for the "consolidated" elastic modulus of the fourth order (see above).

In Fig. 1, as an example, we represent the stress-strain data obtained in Ref. 26 by MD simulations (squares) and by DFT method (diamonds) for armchair configuration of graphene edges. Our fittings of these data using Eq. (43) are shown in Fig. 1 by corresponding curves with parameters: $C_{11}(0) = 328$ N/m, $C_{111} = -1270$ N/m, $C_{1111} = 960$ N/m (MD) and $C_{11}(0) = 328$ N/m, $C_{111} = -1470$ N/m, $C_{1111} = 960$ N/m (DFT). In addition, in Fig. 1 we show a thick curve constructed by Eq. (43) with the values of the parameters $C_{11}(0) = 328$ N/m, $C_{111} = -1000$ N/m, $C_{1111} = 900$ N/m, which we consider as preferable within the framework of our theory; this curve also falls within the error range (dashed lines) of the experiment [23]. Besides, in Fig. 1 we also show the DFT data [24] for zigzag configuration (circles); our fitting of these data by Eq. (43) led to the following values: $C_{11}(0) = 350$ N/m, $C_{111} = -2450$ N/m, $C_{1111} = 7640$ N/m.[6]

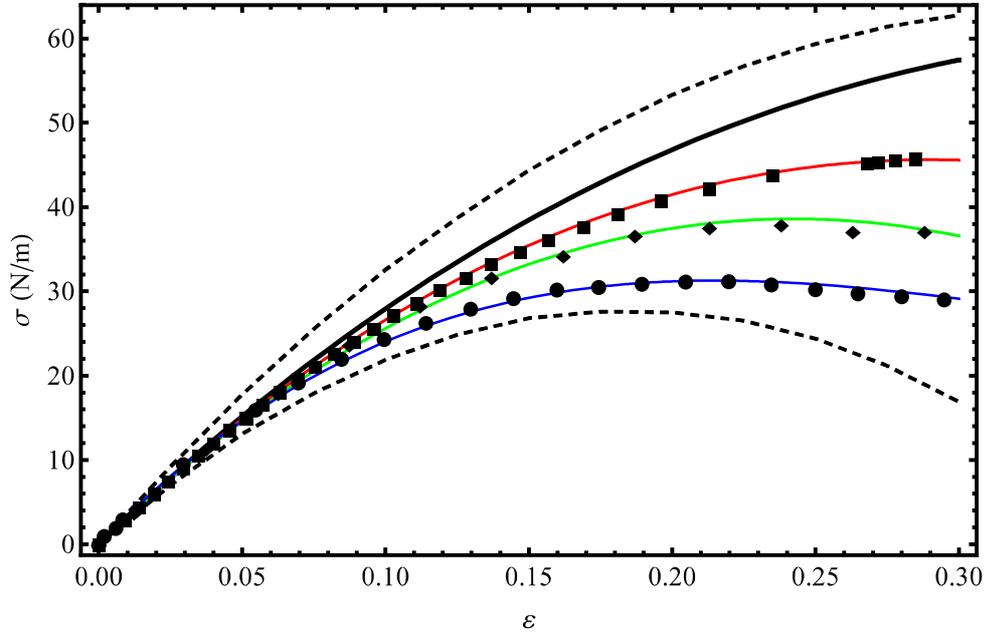

FIG. 1. Stress-strain data for graphene extracted from literature sources. The results obtained in Ref. 26 for armchair configuration of graphene edges are shown by squares (MD simulations) and by diamonds (DFT method). Our processing of these data using Eq. (43) is shown by the corresponding solid lines (the parameter values are in the text). The thick solid curve is constructed according to Eq. (43) with the following parameter values: $C_{11}(0) = 328$ N/m, $C_{111} = -1000$ N/m, $C_{1111} = 900$ N/m. DFT data from [24] are given by circles; their fitting according to Eq. (43) is shown by the corresponding curve (the parameter values are in the text). The dashed lines represent the boundaries of the experimental error [23] (cf. [22]).

---

[6] To approximate the DFT stress-strain data for graphene, in [24] it was used a polynomial containing, besides the terms (43), an additional term of the form $C_{11111}\varepsilon^4/24$. As a result, in [24] it was found: $C_{111} \approx -2800$ N/m, $C_{1111} \approx 13000$ N/m, $C_{11111} \approx -31000$ N/m. Our approximation of the data [24] by a polynomial (43) leads to the value of $C_{1111}$, almost half as much as in [24]. As for the problem to which the present work is devoted, the adding in the Hamiltonian (2) of terms with obviously negative elastic constants of the fifth order leads to the thermodynamic instability of the state of graphene with respect to the large strains. This issue will be discussed below.

Thus, the processing of the simulation stress-strain dependences for graphene (Fig. 1) shows that the data of Ref. 26 are closer to what is predicted by our theory than the data of [24]. Therefore, it can be expected that the estimates of the elastic parameters of graphene at the level of $C_{11}(0) \approx 330$ N/m, $C_{111} \approx -1000$ N/m, $C_{1111} \approx 900$ N/m, can already be used as a basis for describing the phenomenon of graphene melting according to the scheme proposed in the present work. Note that with these values of elastic moduli, using formula (32), we obtain $\eta(0) \approx 0.788$. In addition, taking these estimates, using the above values of the remaining parameters, and setting $T_0 = 1000$ K, we find from expressions (5), (7), (30): $B(T_m) \approx 0.448$ N/m, $B_0(T_m) \approx 578$ N/m. This implies the following estimate: $B(T_m)/B_0(T_m) \approx 7.76 \cdot 10^{-4}$, which justifies the inequality used above.

As already mentioned, from the self-consistent equation (36) at any temperature, one can find a "regular" solution $Q^{(R)} \ll 1$, and then return to the out-of-plane quantity by formula (31):

$$\sqrt{\langle (\nabla w^{(R)})^2 \rangle_w} \equiv \sqrt{\frac{3|C_{111}|}{2C_{1111}} Q^{(R)}} \:. \tag{44}$$

The temperature dependence of this quantity at the above values of parameters is shown in Fig. 2 by curve 1(R) (blue). On the other hand, at $T \geq T_m$, Eq. (36) contains the aforementioned "special" solution $Q^{(S)} \sim 1$, defining another out-of-plane quantity

$$\sqrt{\langle (\nabla w^{(S)})^2 \rangle_w} \equiv \sqrt{\frac{3|C_{111}|}{2C_{1111}} Q^{(S)}} \:. \tag{45}$$

In Fig. 2, curve 1(S) (blue) represents the temperature dependence of the quantity (45) for the above values of the parameters. In addition, curve 2 (red) in Fig. 2 shows the temperature dependence of the in-plane quantity $\sqrt{\langle \partial_\alpha u_\beta \partial_\alpha u_\beta \rangle_u}$ derived by Eq. (7). By the arrow in Fig. 2 we mark a jump of the out-of-plane-quantity into the state with abnormally large (determined by the value $Q_m^{(S)} \approx 0.5$) fluctuations at the melting temperature $T_m = 4950$ K.

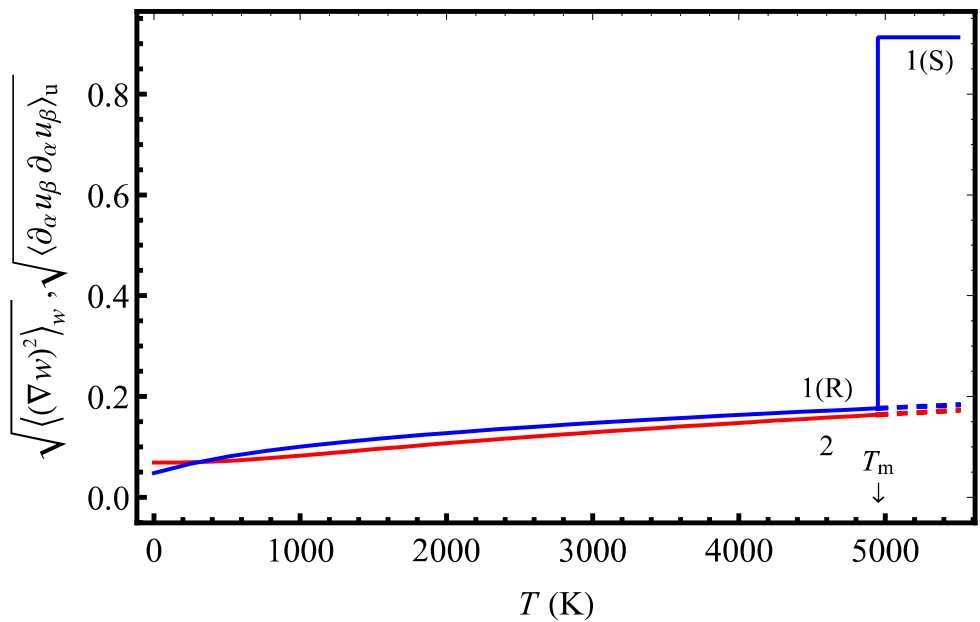

FIG. 2. Temperature dependences of the out-of-plane quantities derived by Eq. (44) [curve 1(R), blue] and by Eq. (45) [curve 1(S), blue]. The temperature-dependent in-plane quantity according to Eq. (7) is shown by curve 2 (red). When building these curves we used the following values of elastic parameters: $C_{11}(0) = 328$ N/m, $C_{111} = -1000$ N/m, $C_{1111} = 900$ N/m. The arrow indicates the melting temperature of graphene $T_m$, at which the jump of the out-of-plane quantity occurs into the state with abnormally large fluctuations. Continuations of the "regular" out-of-plane quantity (44) and the in-plane quantity (7) into the metastable region (i.e. above $T_m$) are indicated by dotted lines.

The analysis shows that the "regular" solution of Eq. (36) at temperature $T_m$ has the value $Q_m^{(R)} \approx 0.017$, whereas the corresponding value of the "pseudo-harmonic" bending modulus of graphene turns out to be small, but quite finite: $B(Q_m^{(R)};T_m) \approx 7.93$ N/m, so that $B(Q_m^{(R)};T_m)/B_0(T_m) \approx 1.37 \cdot 10^{-2}$. Considering now Eq. (44) at $T = T_m$, substituting in it the found value $Q_m^{(R)}$ and the above numerical estimates for anharmonic elastic moduli, we arrive at the following definition of the out-of-plane "Lindemann parameter" for graphene:

$$L_w \equiv \sqrt{\left\langle (\nabla w_m^{(R)})^2 \right\rangle_w} = \sqrt{\frac{3|C_{111}|}{2C_{1111}}Q_m^{(R)}} \approx 0.176. \qquad (46)$$

The dimensionless quantity $L_w$ may be considered as a quantitative characteristic of the melting phenomenon of graphene, similar to the well-known phenomenological melting criterion of Lindemann [2,3]. A close physical connection of $L_w$ with the usual definition of the Lindemann parameter [2] is evident from the fact that $\sqrt{\left\langle (\nabla w_m^{(R)})^2 \right\rangle_w}$ has the meaning of averaged relative "amplitude" of bending thermal vibrations of graphene atoms at the threshold of the appearance of the "special" solution of the self-consistent equation (36). Therefore, the estimate given in (46) even by the numerical value (close to typical values $\approx 0.25$ for 3D solids [2]) can be considered as a direct analog of the Lindemann criterion with respect to the graphene melting.

In addition to (46), it is easy to give an estimate of the averaged relative "amplitude" of thermal vibrations of atoms in the graphene plane at the melting threshold, by entering of the in-plane analog of the "Lindemann parameter":

$$L_u \equiv \sqrt{\left\langle \partial_\alpha u_{\beta m} \partial_\alpha u_{\beta m} \right\rangle_u} \approx 0.164. \qquad (47)$$

We note the proximity of the estimates for $L_u$ and $L_w$ and emphasize once again that for the consistent analysis of the melting phenomenon of graphene, it is urgent to take into account third- and fourth-order anharmonic invariants for the strain tensor in the elastic Hamiltonian [see (2)].

Let us now dwell on one essential point. As is known [4], the important characteristics of a phase transition are the conditions that the thermodynamic functions of the coexisting phases must satisfy. For example, at the melting point of an ordinary 3D solid, its free energy (specific) must coincide with the free energy of the resulting liquid, but the entropy, expressed through the temperature derivative of the free energy, will experience a jump. Let us show that, according to the theory of 2D melting under consideration, already the graphene free energy itself (or rather, its bending part) should experience a jump at the transition point. We will be based on the following expression for the dimensionless free energy of bending vibrations of graphene per cell [see Eq. (19) from Ref. 15]:

$$f_w[\xi_w;\xi_B(Q;T)] \equiv \frac{1}{\xi_w^2}\int_0^{\xi_w} d\xi \ln\left\{1-\exp\left[-\sqrt{\xi^2+\xi_B(Q;T)\xi}\right]\right\} \qquad (48)$$

where the notations are entered:

$$\xi_w \equiv \frac{\theta_w}{T}, \quad \xi_B(Q;T) \equiv \frac{\hbar}{T}\sqrt{\frac{\rho}{\kappa}}s_B^2(Q;T). \qquad (49)$$

By making calculations using formula (48) with the above values of parameters, it can be verified that the free energy corresponding to the "special" root ($f_w^{(S)}$ = –4.812) already at the threshold of its appearance is less than the free energy corresponding to the "regular" root ($f_w^{(R)}$ = –4.518). This means that the transition of graphene to the "melting" state is energetically favorable, although a complete analogy with the thermodynamics of melting of 3D crystals is excluded (see Introduction and Section V).

We now turn to Eq. (32) and give an expression that determines the temperature dependence of the second-order elastic modulus (see Appendix B):

$$C_{11}(T) = C_{11}(0) - C_{11}^{(T)}\langle\partial_\alpha u_\beta \partial_\alpha u_\beta\rangle_{\mathbf{u}}. \qquad (50)$$

Here $C_{11}(0) \equiv \lambda + 2\mu$, and the symbols $\lambda$ and $\mu$, as before, mean the 2D Lamé coefficients of graphene at $T = 0$; below, for the average in (50), we only need to use the classical expression (9). In addition, in (50) it is indicated:

$$C_{11}^{(T)} \equiv |C_{111}| - \frac{1}{2}|C_{112}| - \frac{1}{12}\left(\frac{23}{12}a_4 + \frac{9}{4}b_4 + \frac{7}{3}c_4 + 3d_4\right). \qquad (51)$$

Immediately, we note that by the "minus" sign in the right-hand side of (50), we specially emphasized the fact that $C_{11}(T)$ falls with temperature, as it follows, for example, from [17,35]. Let us set $C_{1111} \equiv a_4+b_4+c_4+d_4 = 900$ N/m, $C_{111} = -1000$ N/m, and $C_{112} \approx C_{111}/5$ [36]. Then, estimating the expression in parentheses (51) as $\approx 2C_{1111}$, we find $C_{11}^{(T)} \approx 750$ N/m. Further, from expressions (50) and (9) in the limit of high temperatures, we obtain:

$$C_{11}(T) = (\lambda+2\mu)\left(1-\frac{T}{T_\infty}\right), \qquad (52)$$

where a certain characteristic temperature appears:

$$T_\infty \equiv \frac{m\mu(\lambda+2\mu)^2}{\rho(\lambda+3\mu)C_{11}^{(T)}}. \qquad (53)$$

Substituting into (53) the well-known [15] numerical values of the graphene parameters: $\rho = 7.6\cdot10^{-8}$ g/cm$^2$, $\mu \approx 3\lambda \approx 9$ eV/Å$^2$, $m = 2\cdot10^{-23}$ g and using the above estimate of $C_{11}^{(T)}$, we find for the characteristic temperature the value $T_\infty \approx 82000$ K. Processing the results of numerical calculations of the graphene elastic characteristics [35] leads to the values $T_\infty \approx 40000 - 53000$ K.[7]

It is useful to note that in energy units, the given estimates for $T_\infty$ are in order of magnitude consistent with the formation energy $E_{S-W}$ of the Stone-Wales defect in graphene. Really, besides the value $E_{S-W} \approx 4.6$ eV [9,10] indicated in the Introduction, in the literature one can find calculated values from 4.8 eV and 5.2 eV [37] to 5.8 eV [38] and even 6.0 eV [39]. It should be borne in mind that the estimation of $T_\infty$ should rather be associated not with $E_{S-W}$, but with the height of the barrier, which separates the ground state of graphene and the state in the presence of the Stone-Wales defect. Taking into account the estimate for this barrier height $\approx 9 - 10$ eV [37], we come to the conclusion that it correlates well with the found above value $T_\infty$. Thus, we see that the proposed model, which considers the fluctuation vibrations of the crystal lattice as the resulting cause of its melting [2,3], does not contradict the "defect" model of melting [3,6,8] (for more details, see Section V).

Note that if we use expression (32), substitute (52) into it at $T = T_m$ and take into account that $\eta(T_m) = 3/4$ (we neglect the term with $B(T_m)/B_0(T_m) \ll 1$), then we obtain the relation:

$$\frac{T_m}{T_\infty} = 1 - \frac{3}{4\eta(0)}. \qquad (54)$$

This relation directly expresses the graphene melting temperature $T_m$ through its characteristic temperature $T_\infty$ and the dimensionless combination of elastic moduli of second, third, and fourth orders at zero temperature (of course, it should be $\eta(0) > 3/4$).

TABLE I. The "controlling" parameter at zero temperature, the characteristic temperature, the melting temperature, the out-of-plane and the in-plane "Lindemann parameters" in dependence of the value of the graphene fourth-order elastic constant.

| $C_{1111}$ (N/m) | $\eta(0)$ | $T_\infty$ (K) | $T_m$ (K) | $L_w$ | $L_u$ |
|---|---|---|---|---|---|
| 890 | 0.779 | 81560 | 4030 | 0.164 | 0.148 |
| 895 | 0.784 | 81650 | 4490 | 0.170 | 0.156 |
| 901 | 0.789 | 81760 | 5040 | 0.178 | 0.165 |
| 906 | 0.793 | 81850 | 5490 | 0.183 | 0.172 |
| 912 | 0.799 | 81960 | 6030 | 0.190 | 0.180 |

Note the sharp dependence of $T_m$ on the difference $[\eta(0) - 3/4]$ (see Table I). This means that in order to obtain reliable information about $T_m$, it is necessary to know as accurately as possible the value of $\eta(0)$ [or the value of $C_{1111}$ for given values of $C_{11}(0)$ and $C_{111}$]. This conclusion is in qualitative agreement with the statement [1] that to accurately predict the melting point (of silicon [1]), it is necessary to know the free energies of the solid and liquid phases with *very high precision*.

## V. DISCUSSION AND CONCLUSIONS

The results obtained in this work allow us to associate the melting phenomenon of a quasi-2D graphene crystal with the temperature transformation of an anomalously soft bending mode, whose "sound" dispersion at small wave numbers is entirely determined by fluctuation effects in the ensemble of in-plane oscillations [13]. Although the melting mechanism itself is determined to a decisive degree by the anharmonic properties of the material, however, precisely due to the natural softening of second-order elastic moduli, graphene at a certain (rather high) temperature can enter a new phase, which in principle does not exist at lower temperatures. This transition is of threshold-type: it is characterized by a sharp increase in the root-mean-square deviations of the graphene surface from the flat. At the same time, the new phase ("melt") already at the point of its origin turns out to be energetically more favorable than the initial – crystalline – phase.

---

[7] Similar processing of the data [17] gives a much smaller value: $T_\infty \approx 16000$ K.

It is necessary to immediately emphasize the difference between the proposed approach and mechanical single-phase melting models [3] (for example, in the old paper [34], the melting temperature was determined by the condition that the shear modulus of a crystal vanishes). In the proposed approach, the melting point of graphene is associated with the sudden appearance of a *new* solution of the self-consistent equation for the root-mean-square fluctuations of the bending strain tensor, and not with some limit condition on the regular – "crystalline" – solution. It is interesting, however, that some effective construction – the "pseudo-harmonic" bending modulus of graphene – still turns out to be very close to zero at the melting point. The melting characteristics of graphene themselves (the melting temperature and the value of the Lindemann constant) naturally depend on its elastic moduli not only of the second, but also of the third and fourth orders. In this case, the analog of the Lindemann constant for graphene calculated in this work turns out to be numerically close to the empirical value ($\approx 0.25$) typical for 3D crystals [2]. It is important to note that all the parameters of the theory, in principle, can be found independently (both from real experiments and using numerical simulation of the elastic characteristics of graphene). Given the fact that the theoretical criterion for the graphene melting includes both harmonic and anharmonic elastic moduli [see (46)], it can be argued, that just through the dimensionless combination of these moduli should be expressed the empirical numerical coefficient, which is present in the purely "harmonic" 3D formula of Lindemann [2].

We pay special attention to the following circumstance. As is well known (see, for example, [3]), along with the so-called "oscillatory" models of melting (on which, in fact, the Lindemann formula is based), there are also approaches in which the melting mechanism is associated with the increasing role of intrinsic crystal defects when approaching the melting point. In our approach, we nowhere explicitly used the concept of crystal defects. It is significant, however, that the very form of Hamiltonian (11) indicates the crucial role of multi-phonon processes in the occurrence of conditions for the transition of graphene into a phase with abnormally large fluctuations of the bending component of the strain tensor. Since the formation of defects (in the case of graphene, this may be the already mentioned Stone-Wales defects [9,10]) is due to precisely multi-phonon processes, the theory proposed in this paper is naturally linked with ideas about the melting of a crystal as the result of its "softening" due to the thermal creation of defects [3,6,8].

Note also that although the paper [24] is so far the only source of information on the numerical values of higher order elastic moduli of graphene (up to the fifth inclusive!), it is hardly possible to directly use these values for our purposes. The point is that in order to coordinate the results of the DFT calculations with the concepts of the theory of elasticity, in [24] it was necessary to introduce into the strain energy the terms of the fifth order with the elastic constants $\approx -32000$ N/m. Meanwhile, the break of expression for the elastic energy on the term with negative elastic constants leads to physically unsatisfactory results, since it allows the formal divergence of the components of the strain tensor as a condition of "minimum" of the strain free energy. To eliminate such a divergence, in the expression for the elastic energy one should take into account the sixth-order terms with *positive* coefficients, which would lead to unnecessarily cumbersome mathematical operations.

The natural way to build a theory, necessarily satisfying the physical requirements of the finiteness of the results and quite mathematically acceptable, is the break of the expansion of the elastic energy density on the fourth-order term over the strain tensor. Thus, we return to Hamiltonian (2) and to all the results obtained above in Sections II–IV. In this case, however, it is necessary to select a suitable fourth-order polynomial for the graphene elastic energy density in order to describe quantitatively the DFT calculation of the stress-strain curves obtained in [24]. Thus, the analysis of the DFT data [24] by formula (43), based on such expansion of the elastic energy density, allows to restore the anharmonic elastic constants with satisfactory accuracy: $C_{111} = -2450$ N/m, $C_{1111} = 7640$ N/m (cf. Fig. 1). The latter value is almost twice as small as $C_{1111} \approx C_{2222} \approx 13000$ N/m, obtained in [24] when taking into account also terms of the fifth-order on strain tensor in the elastic energy density. Interestingly, substituting $C_{111} = -2450$ N/m, $C_{1111} = 7640$ N/m together with $C_{11}(0) = 350$ N/m [24] into formula (32), we come to the value $\eta(0) \approx 1.2 > 3/4$ (cf. the condition at the end of the previous section). This means that, in principle, the results of [24] can be agreed (at least on a qualitative level) with the main conclusions of our theory of graphene melting.

We list the principal results following from our study of 2D melting:

1. Melting should be regarded as a fundamentally non-linear phenomenon, since arbitrarily large displacements of atoms from the initial positions are permissible in the disordered phase. Therefore, it seems impossible to create a consistent theory of melting based on a purely harmonic approximation for a crystal (for example, from the requirement that one of its harmonic elastic moduli vanishes at the melting point [34]). Using the example of graphene as quasi-2D crystal with the bending degree of freedom, we substantiated this statement by constructing a self-consistent theory taking into account third- and fourth-order anharmonic effects. In this case, the decisive fact was that the third order elastic constants for graphene are *negative*;
2. The melting point of graphene is determined by the condition of its jump-like transition into a "special" phase, characterized by abnormally large mean-square bending fluctuations of the strain tensor. On the other hand, a certain effective construction, the "pseudo-harmonic" bending modulus of graphene, reaches an unusually small value at the transition point, which signals the phenomenon of melting. Below the melting point, the theory does not predict any premelting effects, i.e. signs of the existence of a "special" phase ("melt"), even as metastable one;
3. The presence of anharmonic terms in the elastic Hamiltonian of graphene is equivalent to taking into account multi-phonon processes that can lead to the appearance of structural defects. Therefore, the approach proposed in the present paper can be considered as a theoretical alternative to the model based on the ideas about the main role of intrinsic defects in crystal melting;
4. The most important role of the "soft" bending mode in the mechanism of graphene melting is evidenced by the fact that while ignoring the bending degree of freedom, MD simulations [20] lead to a much overestimated (almost twice compared with the generally accepted value ≈ 4500 K [6,8]) melting temperature of graphene. On the contrary, taking into account the bending mode, we can construct a self-consistent theory of 2D melting and obtain a realistic estimate of the melting temperature, as well as find the value of a parameter similar to the well-known 3D Lindemann parameter.

Note that the proposed theory of melting (with specific refinements) can be applied to related 2D crystals: silicene, germanene, etc. Moreover, the conceptual statements of the theory formulated in the present work, in principle, could be extended to 3D crystals, as well. As an important suggestive consideration, we note a well-known experimental fact [40,41] (see also [23]), according to which third-order elastic moduli in solids, as a rule, turn out to be *negative*. In the light of the approach outlined above, this fact is of exceptional importance for the very possibility to realize the condition of a jump-like appearance of a special solution of some self-consistent equation (in the case of graphene, this is the equation for the root-mean-square fluctuations of the "bending" strain tensor). However, it is necessary to make a substantial remark. In contrast to the quasi-2D case where the elastic Hamiltonian can contain only even powers of the derivatives of the out-of-plane displacements, the Hamiltonian of 3D solid can include invariants composed of any powers of spatial derivatives of the elastic displacements. Therefore, when constructing a theory of melting for 3D crystals, these circumstances should be taken into account.

Finally, we draw attention to the next question. The self-consistent theory proposed above made it possible to express the fluctuation characteristics of graphene at temperatures up to the melting point through the "harmonic" (temperature-dependent) modulus, as well as third-order and fourth-order elastic constants. Unfortunately, to date, information on fourth-order elastic moduli is practically absent in the literature devoted to graphene (the exception is work [24], but with the remarks noted above); there is also no consensus regarding the values of the third order elastic constants. In this regard, further work on the study of the anharmonic elastic characteristics of graphene and similar 2D compounds is of utmost importance.

## REFERENCES

[1] F. Dorner, Z. Sukurma, Ch. Dellago, and G. Kresse, Phys. Rev. Lett. **121**, 195701 (2018).
[2] J. M. Ziman, *Principles of the theory of solids* (at the University Press, Cambridge, 1972).
[3] A. R. Ubbelohde, *Melting and crystal structure* (Clarendon Press, Oxford, 1965).
[4] L. D. Landau, and E. M. Lifshitz, *Statistical Physics* (Pergamon, Oxford, UK, 1980), Part 1.
[5] V. M. Bedanov, G. V. Gadiyak, and Yu. E. Lozovik, Phys. Lett. A **109**, 289 (1985).
[6] K. Zakharchenko, J. H. Los, A. Fasolino, and M. Katsnelson, J. Phys.: Condens. Matter **23**, 202202 (2011).
[7] L. D. Landau, and E. M. Lifschitz, *Theory of Elasticity* (Pergamon Press, Oxford, 1959).
[8] J. H. Los, K. V. Zakharchenko, M. I. Katsnelson, and A. Fasolino, Phys. Rev. B **91**, 045415 (2015).
[9] J. H. Los, L. M. Ghiringhelli, E. J. Meijer, and A. Fasolino, Phys. Rev. B **72**, 214102 (2005).
[10] E. Ganz, A. B. Ganz, L.-M. Yang, and M. Dornfeld, Phys. Chem. Chem. Phys. **19**, 3756 (2017).
[11] J. M. Kosterlitz, and D. J. Thouless, J. Phys. C **6**, 1181 (1973).
[12] D. R. Nelson, and L. Peliti, J. de Physique **48**, 1085 (1987).
[13] V. M. Adamyan, V. N. Bondarev, and V. V. Zavalniuk, Phys. Lett. A **380**, 3732 (2016).
[14] R. Ramírez, E. Chacón, and C. P. Herrero, Phys. Rev. B **93**, 235419 (2016).
[15] V. N. Bondarev, V. M. Adamyan, and V. V. Zavalniuk, Phys. Rev. B **97**, 035426 (2018).
[16] N. Mounet, and N. Marzari, Phys. Rev. B **71**, 205214 (2005).
[17] K. V. Zakharchenko, M. I. Katsnelson, and A. Fasolino, Phys. Rev. Lett. **102**, 046808 (2009).
[18] C. P. Herrero, and R. Ramírez, J. Chem. Phys. **148**, 102302 (2018).
[19] C. P. Herrero, and R. Ramírez, Phys. Rev. B **97**, 195433 (2018).
[20] V. V. Hoang, L. T. C. Tuyen, and T. Q. Dong, Phil. Mag. **96**, 1993 (2016).
[21] E. V. Castro, H. Ochoa, M. I. Katsnelson, R. V. Gorbachev, D. C. Elias, K. S. Novoselov, A. K. Geim, and F. Guinea, Phys. Rev. Lett. **105**, 266601 (2010).
[22] E. Cadelano, P. L. Palla, S. Giordano, and L. Colombo, Phys. Rev. Lett. **102**, 235502 (2009).
[23] C. Lee, X. Wei, J. W. Kysar, and J. Hone, Science **321**, 385 (2008).
[24] X. Wei, B. Fragneaud, Ch. A. Marianetti, and J. W. Kysar, Phys. Rev. B **80**, 205407 (2009).
[25] L. D. Landau, and E. M. Lifshitz, *Fluid Mechanics*, 2$^{nd}$ Edition (Elsevier, Oxford, 1987).
[26] G. Kalosakas, N. N. Lathiotakis, C. Galiotis, and K. Papagelis, J. Appl. Phys. **113**, 134307 (2013).
[27] N. M. Plakida and T. Siklós, Phys. Stat. Sol. **33**, 103 (1969).
[28] R. Peierls, Phys. Rev. **54**, 918 (1938).
[29] S. Tyablikov, *Methods in the Quantum Theory of Magnetism* (Springer Science + Business Media, New York, 1967).

[30] R. P. Feynman, *Statistical Mechanics: A Set of Lectures* (W. A. Benjamin, Inc., Advanced Book Program Reading, Massachusetts, 1972).
[31] M. S. Tsai, AIAA Journal **18**, 1180 (1980).
[32] N. Krylov, and N. N. Bogoliubov, *Introduction to Nonlinear Mechanics* (Princeton University Press, Princeton, N.J., 1947).
[33] N. N. Bogoliubov, and Y. A. Mitropolski, *Asymptotic Methods in the Theory of Nonlinear Oscillations* (Gordon and Breach, New York, 1961).
[34] L. Brillouin, Phys. Rev. **54**, 916 (1938).
[35] T. Shao, B. Wen, R. Melnik, Shan Yao, Y. Kawazoe, and Y. Tian, J. Chem. Phys. **137**, 194901 (2012).
[36] L. Colombo, and S. Giordano, Rep. Prog. Phys. **74**, 116501 (2011).
[37] L. Li, S. Reich, and J. Robertson, Phys. Rev. B **72**, 184109 (2005).
[38] J. Ma, D. Alfè, A. Michaelides, and Enge Wang, Phys. Rev. B **80**, 033407 (2009).
[39] L. G. Zhou, and S.-Q. Shi, Appl. Phys. Lett. **83**, 1222 (2003).
[40] O. H. Nielsen, and R. M. Martin, Phys. Rev. B **32**, 3792 (1985).
[41] M. Lopuszynski, and J. A. Majewski, Phys. Rev. B **76**, 045202 (2007).

## APPENDIX A: FINDING SOME "PSEUDO-HARMONIC" AVERAGES FOR THE BENDING MODE

Here we outline the path for obtaining formula (19), considering the mean $\langle (\nabla w)^6 \rangle_w$ as an example. Let us proceed under the sign of the mean to the 2D Fourier representation (13) and take into account that in the pseudo-harmonic approximation the averaging will occur with the Gaussian distribution function

$$\exp\left[-\frac{1}{T}\sum_{\mathbf{k}}\psi(k)w_{\mathbf{k}}w_{-\mathbf{k}}\right], \tag{A1}$$

where $\psi(k)$ is an amplitude. Then we have:

$$\langle (\nabla w)^6 \rangle_w = \frac{i^6}{S^3}\left\langle \sum_{\mathbf{k}_1,\ldots,\mathbf{k}_6} k_{1\alpha}k_{2\alpha}k_{3\beta}k_{4\beta}k_{5\gamma}k_{6\gamma} w_{\mathbf{k}_1}w_{\mathbf{k}_2}w_{\mathbf{k}_3}w_{\mathbf{k}_4}w_{\mathbf{k}_5}w_{\mathbf{k}_6}\delta_{\mathbf{k}_1+\mathbf{k}_2+\mathbf{k}_3+\mathbf{k}_4+\mathbf{k}_5+\mathbf{k}_6}\right\rangle_w, \tag{A2}$$

where $\delta_{\mathbf{k}_1+\mathbf{k}_2+\mathbf{k}_3+\mathbf{k}_4+\mathbf{k}_5+\mathbf{k}_6}$ is the Kronecker $\delta$-symbol (1 or 0). Eliminating one of the sums with the help of the $\delta$-symbol, one can lead the right-hand side of (A2) to the form:

$$\frac{4}{S^3}\left\langle \sum_{\mathbf{k}_1,\ldots,\mathbf{k}_5} k_{1\gamma}k_{1\alpha}k_{2\alpha}k_{3\beta}k_{4\beta}k_{5\gamma} w_{\mathbf{k}_1}w_{\mathbf{k}_2}w_{\mathbf{k}_3}w_{\mathbf{k}_4}w_{\mathbf{k}_5}w_{-\mathbf{k}_1-\mathbf{k}_2-\mathbf{k}_3-\mathbf{k}_4-\mathbf{k}_5}\right\rangle_w$$
$$+\frac{1}{S^3}\left\langle \sum_{\mathbf{k}_1,\ldots,\mathbf{k}_5} k_5^2 k_{1\alpha}k_{2\alpha}k_{3\beta}k_{4\beta} w_{\mathbf{k}_1}w_{\mathbf{k}_2}w_{\mathbf{k}_3}w_{\mathbf{k}_4}w_{\mathbf{k}_5}w_{-\mathbf{k}_1-\mathbf{k}_2-\mathbf{k}_3-\mathbf{k}_4-\mathbf{k}_5}\right\rangle_w. \tag{A3}$$

Further, we take into account the fact that with Gaussian averaging a nonzero contribution to (A3) will be obtained only in cases where each harmonic will be in even powers. This allows rewriting (A3) in the form:

$$\frac{4}{S^3}\left\langle \sum_{\mathbf{k}_1,\ldots,\mathbf{k}_5} k_{1\gamma}k_{1\alpha}w_{\mathbf{k}_1}w_{-\mathbf{k}_1}k_{2\alpha}k_{3\beta}k_{4\beta}k_{5\gamma}\delta_{\mathbf{k}_2+\mathbf{k}_3+\mathbf{k}_4+\mathbf{k}_5}w_{\mathbf{k}_2}w_{\mathbf{k}_3}w_{\mathbf{k}_4}w_{\mathbf{k}_5} \right\rangle_w$$

$$+\frac{1}{S^3}\left\langle \sum_{\mathbf{k}_1,\ldots,\mathbf{k}_5} k_5^2 w_{\mathbf{k}_5}w_{-\mathbf{k}_5}k_{1\alpha}k_{2\alpha}k_{3\beta}k_{4\beta}\delta_{\mathbf{k}_1+\mathbf{k}_2+\mathbf{k}_3+\mathbf{k}_4}w_{\mathbf{k}_1}w_{\mathbf{k}_2}w_{\mathbf{k}_3}w_{\mathbf{k}_4} \right\rangle_w. \quad (A4)$$

Noticing now that $\left\langle k_{1\gamma}k_{1\alpha}w_{\mathbf{k}_1}w_{-\mathbf{k}_1}\right\rangle_w = (1/2)\delta_{\gamma\alpha}\left\langle k_1^2 |w_{\mathbf{k}_1}|^2 \right\rangle_w$, we obtain from (A4):

$$\left\langle (\nabla w)^6 \right\rangle_w = 3\frac{1}{S}\left\langle \sum_{\mathbf{k}_1} k_1^2 |w_{\mathbf{k}_1}|^2 \right\rangle_w \frac{1}{S^2}\left\langle \sum_{\mathbf{k}_2,\ldots,\mathbf{k}_5} k_{2\alpha}k_{3\alpha}k_{4\beta}k_{5\beta}\delta_{\mathbf{k}_2+\mathbf{k}_3+\mathbf{k}_4+\mathbf{k}_5}w_{\mathbf{k}_2}w_{\mathbf{k}_3}w_{\mathbf{k}_4}w_{\mathbf{k}_5} \right\rangle_w$$

$$= 3\left\langle (\nabla w)^2 \right\rangle_w \left\langle (\nabla w)^4 \right\rangle_w = 3\left\langle (\nabla w)^2 \right\rangle_w \cdot 2\left\langle (\nabla w)^2 \right\rangle_w^2 = 3!\left\langle (\nabla w)^2 \right\rangle_w^3. \quad (A5)$$

It should, however, be noted that formula (A5) is strictly correct only in the macroscopic limit ($S\to\infty$). Indeed, in the derivation of (A5), we discarded averages like

$$\frac{1}{S^3}\left\langle \sum_{\mathbf{k}_1} k_1^2 |w_{\mathbf{k}_1}|^2 \right\rangle_w \left\langle \sum_{\mathbf{k}_2} k_2^4 |w_{\mathbf{k}_2}|^4 \right\rangle_w,$$

the statistical weight of which, by virtue of (A1), will be negligible (proportional to $1/S$) compared with the terms presented in (A4).

The general expression (19) can be obtained by analogy with (A5), using the method of mathematical induction.

## APPENDIX B: CALCULATING THE TEMPERATURE DEPENDENCE OF $C_{11}(T)$

To calculate the temperature dependence of the "harmonic" elastic modulus $C_{11}(T)$ [see (11)] we proceed from the general expression (2), in which we specially select the terms, resulting in contributions $\sim (\nabla w)^4$ under the integral. For convenience, we first break (1) into two contributions (in-plane and out-of-plane ones):

$$\varepsilon_{\alpha\beta} \equiv u_{\alpha\beta} + w_{\alpha\beta}, \quad u_{\alpha\beta} \equiv \frac{1}{2}(\partial_\alpha u_\beta + \partial_\beta u_\alpha + \partial_\alpha u_\gamma \partial_\beta u_\gamma), \quad w_{\alpha\beta} \equiv \frac{1}{2}\partial_\alpha w \partial_\beta w. \quad (B1)$$

The resulting dependence $C_{11}(T)$ is obtained both from terms of type $u_{\alpha\beta}w_{\alpha\beta}w_{\gamma\gamma}$ and from terms of type $u_{\alpha\beta}u_{\beta\gamma}w_{\alpha\delta}w_{\delta\gamma}$.

Let us first write the contributions of interest to us, derived from the terms of the second and third orders in $\varepsilon_{\alpha\beta}$ under the integral in (2):

$$\left(\frac{\lambda}{2}+\mu\right)w_{\alpha\alpha}w_{\beta\beta} + \frac{C_{111}-C_{112}}{4}(2u_{\alpha\beta}w_{\alpha\beta}w_{\gamma\gamma} + u_{\gamma\gamma}w_{\alpha\beta}w_{\alpha\beta})$$

$$+\frac{3C_{112}-C_{111}}{12}(2u_{\alpha\alpha}w_{\beta\beta}w_{\gamma\gamma} + u_{\gamma\gamma}w_{\alpha\alpha}w_{\beta\beta}). \quad (B2)$$

From the fourth order terms on $\varepsilon_{\alpha\beta}$ under the integral in (2) we get the following construction:

$$\frac{1}{4!}\{a_4[2u_{\alpha\beta}u_{\beta\gamma}w_{\alpha\delta}w_{\delta\gamma} + (u_{\alpha\beta}w_{\beta\gamma} + u_{\beta\gamma}w_{\alpha\beta})(u_{\gamma\delta}w_{\delta\alpha} + u_{\delta\alpha}w_{\gamma\delta})]$$
$$+ b_4(u_{\alpha\beta}u_{\alpha\beta}w_{\gamma\gamma}w_{\delta\delta} + 4u_{\alpha\beta}u_{\gamma\gamma}w_{\alpha\beta}w_{\delta\delta} + u_{\delta\delta}u_{\gamma\gamma}w_{\alpha\beta}w_{\alpha\beta})$$
$$+ c_4(2u_{\alpha\beta}u_{\alpha\beta}w_{\gamma\delta}w_{\gamma\delta} + 4u_{\alpha\beta}u_{\gamma\delta}w_{\alpha\beta}w_{\gamma\delta}) + 6d_4 u_{\alpha\alpha}u_{\beta\beta}w_{\gamma\gamma}w_{\delta\delta}\}. \quad (B3)$$

The next step consists in averaging expressions (B2) and (B3) over the "fast" in-plane fluctuation oscillations of graphene (cf. [13,15]). It should be noted that non-zero averages in (B2) arise from bilinear contributions contained in $u_{\alpha\beta}$, while in (B3) – from linear contributions contained in each of the two factors of type $u_{\alpha\beta}$. In addition, we will neglect the averages of the type $\langle \partial_\alpha u_\beta \partial_\gamma u_\delta \partial_\zeta u_\eta \partial_\sigma u_\tau \rangle_{\mathbf{u}}$ as quantities of the next order of smallness. Then, collecting the terms resulting from the averaging (B2) and (B3), taking into account the equalities $\langle \partial_\alpha u_\alpha \partial_\beta u_\beta \rangle_{\mathbf{u}} = \langle \partial_\alpha u_\beta \partial_\beta u_\alpha \rangle_{\mathbf{u}} = \langle \partial_\alpha u_\beta \partial_\alpha u_\beta \rangle_{\mathbf{u}}/2$ [13,15] and omitting the intermediate actions, we find the resulting contribution $\sim (\nabla w)^4$ to the "bending" Hamiltonian of graphene (11)

$$\frac{1}{8}\left\{\lambda + 2\mu + \frac{1}{2}\left[2C_{111} - C_{112} + \frac{1}{6}\left(\frac{23}{12}a_4 + \frac{9}{4}b_4 + \frac{7}{3}c_4 + 3d_4\right)\right]\langle \partial_\alpha u_\beta \partial_\alpha u_\beta \rangle_{\mathbf{u}}\right\}(\nabla w)^4. \quad (B4)$$

Finally, comparing expression (B4) with the general form of the corresponding contribution to (11), we arrive at formulas (50), (51) of the main text.